# Exploring Higher Education Competencies through Spreadsheet Self-Assessment & Time

[1]Maria Csernoch, [2]Judit T. Kiss, [3]Viktor Takács, [2]Domicián Máté

University of Debrecen

[1]Faculty of Informatics, [2]Faculty of Engineering, [3]Faculty of Economics

csernoch.maria@inf.unideb.hu, tkiss@eng.unideb.hu, takacs.viktor@econ.unideb.hu, mate.domician@eng.unideb.hu

**ABSTRACT**

The present paper aims to explore higher education students' spreadsheet competencies and reliability through self-assessment and real-world problem-solving practices. Digital natives alleged skills and competences allowed us to hypothesize that students perform better in Excel than on paper, but the findings cannot confirm this hypothesis. However, our results indicate that students tend to inaccurately assess their spreadsheet competencies compared to their actual performance in both paper-based and Excel tasks. It has also be found that students need at least twice as much time to achieve the same high scores in the digital environment as they do on paper. The results violated the wildly accepted assumption that digital native students do not need computer science education, since they are born with it. This study highlights the importance of accurate self-assessment in digital skill development and time management within higher education contexts, particularly in technology-driven disciplines.

## 1. INTRODUCTION

When students realize they should solve spreadsheet tasks on paper – creating formulas and telling what formulas do – their reactions express that they are asked the impossible. The loudest and bravest keep convincing the teacher that completing these requirements is impossible, old-fashioned, and useless. "I cannot do it on paper." "Excel tasks cannot be solved on paper" "I can do it in Excel." "I can find it in the help." "I can search the internet." "I'll ask ChatGPT." The question is who is right: those who claim that spreadsheet problems can only be solved in spreadsheet applications or those who claim that handling spreadsheet data is much more demanding and complex and requires a broader range of knowledge, skill, and competencies than handling spreadsheet applications.

### 1.1. Problem-solving approaches

We all remember Panko's presentation and paper entitled 'The Cognitive Science of Spreadsheet Errors: Why Thinking is Bad.' Panko provided a detailed theoretical background on how knowledge is built up, how cognitive load takes its toll (Sweller et al., 2011), and how fast and slow thinking work (Kahneman, 2011) (calling schemata vs. solving unique problems and building up schemata, respectively). He concluded that the primary source of spreadsheet errors is 'thinking', more precisely 'slow thinking' (Panko, 2013, Kahneman, 2011). Later, Csernoch and her research fellows (Biró & Csernoch, 2014; Csernoch & Biró 2015; Csernoch 2017; Nagy & Csernoch, 2023) found that Panko's results can further be tuned, considering end-users activities carried out in spreadsheet environment. It is concluded that not thinking is bad, but the circumstances where thinking is applied. To find proof for their hypothesis, the typology of computer problem solving approaches (TCPSA) was set up. It defines two hypernyms and two and three hyponyms for categorizing and explaining computer problem-solving methods (Figure 1).

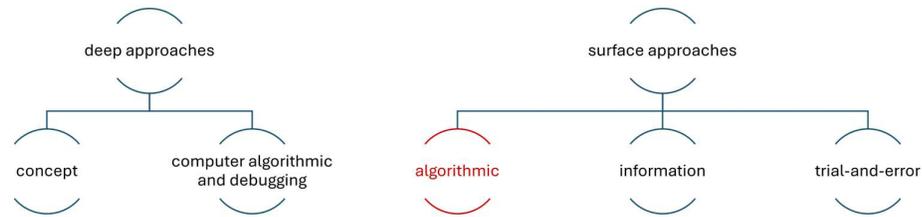

Figure 1. The typology of computer problem-solving approaches.

Based on the typology of computer problem-solving approaches, it is revealed that the right proportion of fast and slow thinking must be found, and these two must be applied to the requirements of the task to be solved.

TCPSA reveals four slow thinking problem-solving approaches – concept, computer algorithmic and debugging, information, and trial-and-error-based – while only one – algorithmic – which approach requires fast thinking (Figure 1). To apply fast thinking, firm schemata must be built up in long-term memory and called when the problem requires it. In any other case, slow thinking must be applied to a problem, which is an erroneous process. In deep approaches, slow thinking is required and accepted. However, in surface approaches, where primarily surface navigation takes place, slow thinking causes errors. This can be explained by the nature of both information and trial-and-error-based approaches since they are not real problem-solving methods but imitated. In these approaches, end-users apply slow thinking navigation and searching methods without being aware of the target conditions of the original problem. The difference between the information and the trial-and-error-based approaches is that in the information-based methods, end-users know how to navigate – nor rarely accompanied by flash-speed mouse and keyboard movements –, while in the trial-and-error methods, they do not have any idea, they keep moving around aimlessly.

Panko also stated that "…spreadsheet developers and corporations are highly overconfident in the accuracy of their spreadsheets.", "…we [humans] are aware of very few of the errors we make.", "humans cannot be error free no matter how hard they try", and "…our intuition about errors and how we can reduce them is based on appallingly bad knowledge." (Panko, 2015). Considering all these findings, we ventured into an even more swampy territory to discover how tertiary education students in end-user role evaluate their spreadsheet knowledge and competencies and how they can perform both on paper and spreadsheet along with its accompanying environments, including tooltips, error messages, automated corrections, helps, online searches, chatbots.

## 1.2. Revealing problem-solving strategies

Problem-solving strategies are one of the black holes of spreadsheeting. As common phrases suggest (listed some at the beginning of the paper), end-users believe that paper is not for solving spreadsheet-related problems anymore since they can do anything in the application or ask ChatGPT or its companions.

However, several questions arise when high-demanding erroneous documents circulate, causing serious losses in human and machine resources. How effectively and efficiently end-users can work, how they can develop their knowledge, skills and competencies in an all-digital environment, how they can avoid, detect and correct errors, how they can use fast and slow thinking, how they can deal with the cognitive load they are aware of cognitive load and knowledge inventory?

In order to answer some of these questions, our research group found that end-users' problem-solving strategies should be revealed, and based on these results, we would be able to compare various teaching-learning methods and set up criteria for effective and efficient teaching-learning and working strategies.

## 2. RESEARCH GOALS

### 2.1. Goals and target conditions

The present study is part of a more extraordinary project that aims to reveal students' and end-users' problem-solving strategies in spreadsheeting. To achieve our goals, we have developed novel tools that would reveal the results and end-users' activities, which are usually hidden from anyone or stay unnoticed. The essence of these tools is to log end-users' activities (Nagy & Csernoch, 2023). Based on the logged data, we would be able to calculate the time spent on the problem, detect the problem-solving strategies, calculate the entropy (the information content), and the sustainability rate of the selected method (Csernoch et al. 2022, 2023, 2024).

The target conditions of the present study are to reveal how students can evaluate their spreadsheet knowledge, how this evaluation matches their performance in solving spreadsheet problems on paper and spreadsheet environments, and how their paper and spreadsheet results can be compared. The question was whether the Dunning-Kruger Effect – "…incompetent individuals have more difficulty recognizing their true level of ability than do more competent individuals and that a lack of metacognitive skills may underlie this deficiency" – works in this situation (Kruger & Dunning, 1999; Gibbs et al., 2011, 2014, 2017), and how the testing situation changes affect students' self-assessment.

### 2.2. Hypotheses

At the beginning of the test, the informal interviews revealed that most of the tested students seemed confident, especially in their digital skills and knowledge, which is in complete accordance with attitudes assigned to digital natives by Prensky (2001a, 2001b). They expressed their long-term connection with spreadsheeting (Excel) and excitement and willingness to participate in the test. However, we must also mention that there were Bachelor students who, after studying informatics in elementary and high schools for at least five years – according to the National Curriculum of 2012 – claimed that they did not have any spreadsheeting experience. Based on the students' expectations and results published earlier (László et al., 2022; Máté & Darabos, 2017), we set up the following hypotheses.

[H1] Students can create formulas more effectively in Excel than on paper. They generally do better in Excel than on paper.
[H2] Students are less accurate in assessing their spreadsheet knowledge, skills and competences in Excel than in solving spreadsheet problems on paper.

Both hypotheses are set up based on the students' suggestions before starting the testing process.

## 3. RESEARCH METHODS

### 3.1. The sample

The research was carried out at our university, where 173 students of various technology-driven courses were tested (Table 1). For the present study, the level of studies (Bachelor or Master, B or M, respectively) and the nationalities of the students (international or Hungarian, INT or HUN, respectively) served as independent variables. The language used in the international courses is English.

The test was anonymous, but the students' University ID was used to compare the results of the three sections.

|  | International (INT) | Hungarian (HUN) |
|---|---|---|
| Bachelor (B) | 12 | 92 |
| Master (M) | 24 | 45 |

Table 1. Students participating in the study.

The testing was conducted in computer labs where all the students had their workstation or laptop connected to the university network, which students could connect through their University ID. The computer rooms have Hungarian and English versions of MS Office through university licenses.

### 3.2. Sections

The research method detailed in the present paper consists of three phases:

- self-assessment to evaluate students' spreadsheet knowledge and competencies on paper (SAV1–SAV4),
- solving spreadsheet problems on paper (P15 and P17 for Tasks 1–5 and Tasks 1–7, respectively),
- solving spreadsheet problems on a spreadsheet interface.

Time limits were not set; all the students were allowed to work as long as they thought they could add value to their solution.

Dear Student,

The ANLITA (Atomic Natural Language Input Tracker Application) research is designed for logging and measuring the effectiveness of digital documents and problem-solving strategies. For evaluation purposes, please indicate in percentage how you rate your spreadsheet knowledge: 0% – I know nothing, 100% – I am very proficient.

Please, be aware that previously entered values cannot be changed.

| Before seeing the test | After reading the test |
|---|---|
|  |  |

| After finishing the paper test | After finishing the Excel test |
|---|---|
|  |  |

Figure 2. The four phases of the self-assessment test.

The tasks presented in the test (Figure 3) are in complete accordance with the requirements of our digital literacy course books (Varga et al., 2020; Abonyi-Tóth et al., 2021, 2022) and maturation exam (OH, 2020) and cover the related requirements of both ECDL (ICDL) (ICDL, 2023) and MOS exams (MS, 2024). The table is downloaded from the mock exam period of the Hungarian maturation exam in 2004 (OH, 2004).

Figure 3. The paper (left) and Excel (right) versions of the tasks.

One of the significant characteristics of the test is that Tasks 3–5 are built on the same algorithm – conditional calculations – and Tasks 6–7 provided the algorithm (Csapó et al., 2019, 2020; Csernoch et al., 2021; Nagy et al., 2021).

### 3.3. The process, the sequence of testing phases

In the first phase, the students filled in the upper left cell of the self-assessment test (Figure 2) to tell how they evaluate their spreadsheet knowledge, skills, and competencies based on their previous studies and experiences.

In the second phase, the students received the test on a sheet of paper (Figure 3), read the seven tasks, and then filled in the upper right cell of the self-assessment test, still focusing on their spreadsheet knowledge. It was emphasized that not their solution should be evaluated but their spreadsheet knowledge.

In the third phase, the tasks were solved on paper without help. This phase took around 10–15 minutes. After handing in the papers, the third evaluation phase arrived.

In the fourth phase, the students worked in a prepared Excel workbook, and their activities were logged with ANLITA (Atomic Natural Language Input Tracker Application). The output of this phase is three files for each student:

- the modified Excel workbook, which holds the student's solution,
- a text-based log file, which collected all the keyboard and mouse activities during the problem-solving process and
- a video file that records the complete process carried out on the screen.

After finishing the Excel solution, the fourth evaluation took place.

Like the paper solution, students were allowed to work as long as they wished and to use any technical help (tooltip, insert function (*fx*), argument panels, online help, chatbots), including mobile phones (which activities were not recorded). The time of this phase is recorded in the text and the video files, which were approximately 2–3 times longer than on paper (detailed in Section 4.3).

### 3.4. Tools

As mentioned in the previous sections, the testing process required various tools, as presented in Table 2.

| tool | target condition |
|---|---|
| self-assessment sheet with four cells (Figure 2) | to students' spreadsheet knowledge during the testing process |
| the paper test of seven tasks with a sample table and two highlighted cells for variables (Cells G2 and G3) (Figure 3 left) | to answer Tasks 1–5 with spreadsheet formulas and Tasks 6–7 with natural language sentences |
| a PDF file of five tasks, a sample table, two highlighted cells for variables (Cells G2 and G3), four highlighted for Answers 1, 3, 4, and 5, and one highlighted column for Answer 2. (Figure 3 right) | to guide students during the problem solving process |
| *countries.xlsx* (Figure 4) | to solve Tasks 1–5 in the indicated cells and column<br>to save the file with the original name |
| ANLITA | to log the keyboard and the mouse activities in a text file<br>to record the problem-solving process carried out on the screen |
| university's online drives, learning frame system, and private e-mail accounts in case of emergency | to share the original files, the modified *countries.xlsx* file, and the two log files |
| Excel | to record data, to provide preliminary results |
| SPSS | to do statistical analysis |

Table 2. Tools to carry out the testing process.

In the original *countries.xlsx* file, Cells G2 and G3 were empty, and students were allowed to add any number they thought matched. Furthermore, unlike the sample in Figure 3, the rows were not hidden, and the columns were not adjusted to the content. Students were allowed to hide rows, change column width, and freeze panes based on the sample tables or convert data into table (Figure 5).

Figure 4. The field names and the first five records of the countries.xlsx file.

Figure 5. Two possible arrangements of the original data.

## 4. RESULTS

The participating students' answers to Tasks 1–7 on paper, Tasks 1–5 in Excel, and their self-evaluation SAV1–4 were recorded, compared, and analysed in the following. The data were collected from the answer and the evaluation sheets (paper), the Excel workbook named *countries.xls*x and the text and video (log) files recorded by ANLITA (Nagy & Csernoch, 2023).

### 4.1. Item-points for recognizable pieces

In order to evaluate the students' results, previously published methods were applied (Csapó et al., 2019, 2020; Csernoch et al., 2021; Nagy et al., 2021). The answers were broken down into recognizable items, and all the answers were corrected and evaluated with item points (1 for the correct answer and 0 for the incorrect). When the item points were available, their sum was calculated and converted to a percentage. Along with the item points, it was also recorded whether the students answered with formulas in Tasks 1–5 and with natural language sentences in Tasks 6–7 or with constants.

It was also recorded in Tasks 3–5 whether the students used built-in formulas or tried to build up the algorithms. Two evaluation sheets were set up for the two solutions using these two options.

In the first step, the mean, standard deviation, and normality of the answers were tested. It is found that for Tasks 1–5 both on paper (P15) and in Excel, the distribution of the item points is not normal (Shapiro-Wilk Statistics = 0.975 p = 0.003; Shapiro-Wilk Statistics = 0.960, $p < 0.001$, respectively). The distribution of the results shows several peaks in the histograms, which dwell on the lower half (Figure 6 and Figure 7).

|       | Mean  | Std   |
| ----- | ----- | ----- |
| P15   | 35.44 | 22.42 |
| Excel | 35.13 | 24.41 |

Table 3. Students' results in the test Task 1–5 on paper (P15) and in Excel.

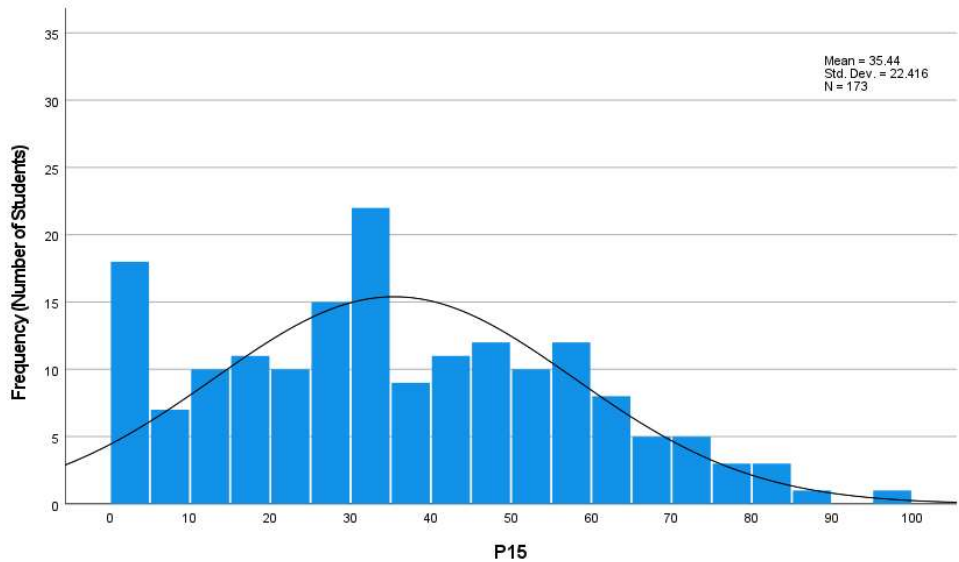

Figure 6. The distribution of the results of Tasks 1–5 on paper.

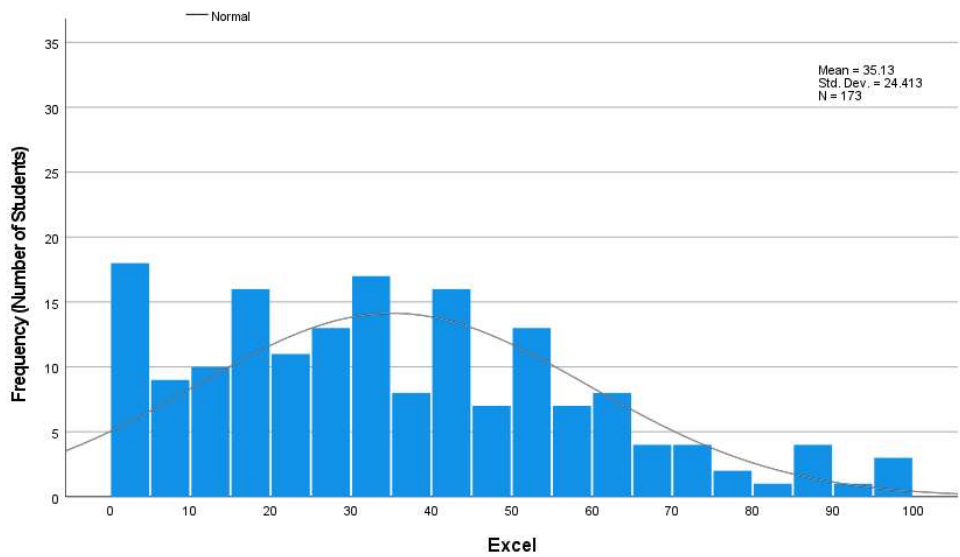

Figure 7. The distribution of the results of Tasks 1–5 in Excel.

Testing for mean difference, the Wilcoxon Signed Ranks Test revealed no significant difference between the paper (P15) and the Excel results (z statistics = −3.78, p = 0.705). The result rejects the first hypothesis [H1]. However, for further results and proofs, we carried out additional statistical analyses (Section 4.3).

### 4.2. Constant answers

Unexpected difficulties made the test more complex than we had planned. Students should have realized that the sample table had 235 records and only considered those presented in the picture (Figure 3 and Figure 8). One of the consequences of this approach was that students tried to answer with constant values or with the imitation of formulas (Figure 8 and Figure 9).

1. What is the capital city of the largest country?

   Algiers

2. What is the population density of each country?

   thousands

3. How many African countries are in the table?

   4

6. What is the result of the following formula?
   {=SUM(IF(B2:B236="Europe",IF(LEFT(A2:A236)="A",1)))}

   3

7. What is the result of the following formula?
   {=SUM((B2:B236="Europe")*(LEFT(A2:A236)="A"))}

   2

---

1. What is the capital city of the largest country?

   Algiers C4 (1)

2. What is the population density of each country?

   E, E236 (2)

3. How many African countries are in the table?

   A4, A4, A7, A235, A236 (4)

6. What is the result of the following formula?
   {=SUM(IF(B2:B236="Europe",IF(LEFT(A2:A236)="A",1)))}

   (Albania + Andorra + Yugoslavia)

7. What is the result of the following formula?
   {=SUM((B2:B236="Europe")*(LEFT(A2:A236)="A"))}

   (Albania

---

1. What is the capital city of the largest country?

   Algiers

2. What is the population density of each country?

   Asia = 27756/607500, Albania = 28749/3565, Algeria = 32278/2381750, American Samoa = 69/199, Andorra = 69/469, Angola = 10389/1246700, Anguilla = 12/102, Yemen = 18821/527970, Yugoslavia = 10657/102350, Zambia = 9759/752614, Zimbabwe = 11877/390580

3. How many African countries are in the table?

   4

3. How many African countries are in the table?

   Four   B4, B7, B235, B236

4. What is the average population of those countries whose surface area is smaller than G2?

   S = AVG E3, E5, E6, E8, E233, E236

5. How many countries have a surface area greater than G3?

   seven = AVERAGE G2, G4, G7, G233, G234 (seven)
                                    ? G235, G236

Figure 8. Various constant-answers on paper.

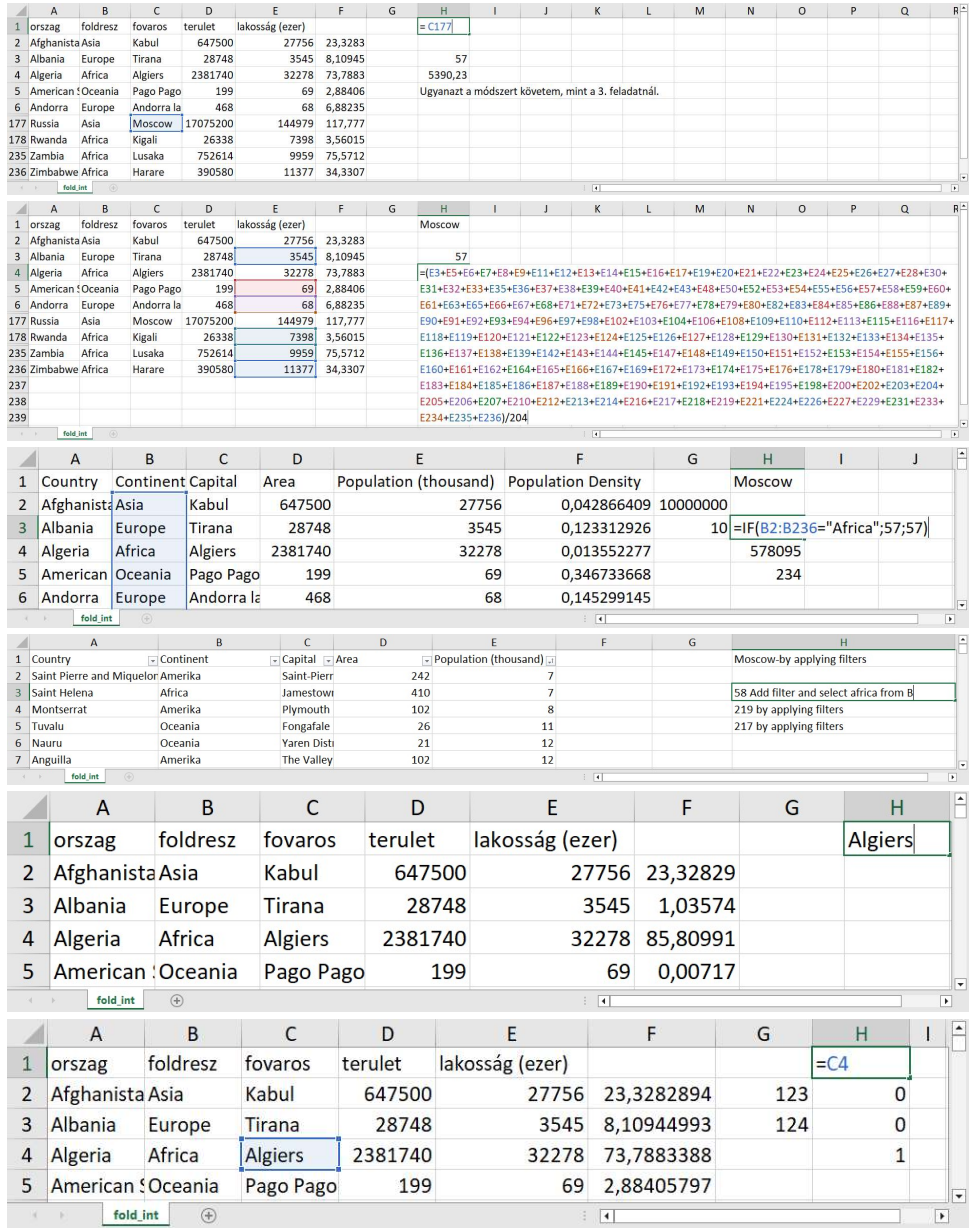

Figure 9. Various constant-answers in Excel.

|  |  | Excel | | paper | |
|---|---|---|---|---|---|
|  | N | NoS | NoC | NoS | NoC |
| INT | 36 | 13 | 30 | 25 | 67 |
| HUN | 137 | 16 | 28 | 11 | 25 |

Table 4. In comparing international and Hungarian students, the number of students (NoS) and constants (NoC) found in the paper and the Excel solutions.

In the comparison of International (INT) and Hungarian (HUN) students, among the 173 students, 29 (16.8%) in Excel and 36 (20.8%) on paper did not understand how to create Excel formulas (Tasks 1–5) or how to express the results of a formula in a natural language sentence (Tasks 6–7). 36.1% of the international and 11.7% of the Hungarian students answered the questions with constants. These students entered 56 in Excel and 92 constant values on paper (Table 4).

In comparing the Bachelor and the Master students, 17.4% of the Master and 16.6% of the Bachelor students entered constants. Details are presented in Table 5.

As long as we wanted to provide as much technical help as necessary to allow flow during the testing, we explained several times that not seeing numbers in G2 and G3 means they are variables. Students can image and add any values suitable for solving the problems. These questions indicate that several students needed help in understanding fundamental spreadsheeting and programming concepts.

|   |   | Excel | | paper | |
|---|---|---|---|---|---|
|   | N | NoS | NoC | NoS | NoC |
| M | 69 | 12 | 31 | 21 | 59 |
| B | 104 | 17 | 27 | 15 | 33 |

Table 5. Number of students (NoS) and constants (NoC) found in the paper and the Excel solutions in comparing Bachelor and Master students.

### 4.3. Solutions and time spent on solutions

We also tried calculating the time spent on the different solutions, paper vs. Excel. As mentioned above, our goal was to recognize problem-solving strategies and patterns. Consequently, the more problems they solved, the more data could be collected. Following this concept, we did not set up a time limit; students were allowed to work as long as they thought they were ready. They generally spent about 10–15 minutes solving the seven tasks on paper and much longer in Excel. Primarily, the international students worked longer in Excel, but their results did not correlate with their working time. During the test, we noticed that students spent most of their time searching both on the computer and mobile devices. However, to recognize patterns, the log files must be analysed, which is the next phase of the research.

Approximately, students spent 2 minutes on each task on paper and 5 minutes in Excel. This means that we should consider the time factor when comparing the results.

It was found that both on paper and in Excel, the Hungarian students' results were higher than those of the international students. On the other hand, the time spent on these tasks was higher for the international students than for the Hungarians.

|   | paper (%) | Excel (%) | time (s) | time (min) |
|---|---|---|---|---|
| INT | 30.31 | 25.10 | 2284 | 38 |
| HUN | 36.40 | 38.16 | 1361 | 23 |

Table 6. Students' results on paper and Excel and the time spent on these solutions in the comparison of international and Hungarian students.

On paper, the results of Master students are higher than those of Bachelor students, while in Excel, it is the other way around. Master students spend more time-solving tasks in Excel than bachelor students.

|   | paper (%) | Excel (%) | time (s) | time (min) |
|---|---|---|---|---|
| M | 37.77 | 34.17 | 1753 | 29 |
| B | 33.38 | 36.28 | 1420 | 24 |

Table 7. Students' results on paper and in Excel and the time spent on these solutions compared to Bachelor and Master students.

As mentioned above, there is no significant difference between the paper and the Excel results, however, we must recognize the time spent on different solutions. Considering the time factor, the students perform significantly better on paper, confirming that students cannot evaluate their spreadsheet knowledge and skills. This result further strengthens the rejection of hypothesis [H1].

It is also found (Mann-Whitney U Test) that there is no significant difference between the paper (P15) and Excel results when Bachelor and the Master students are compared ($Z(P15) = -0.678$, $p = 0.498$, $Z(E) = -1.036$, $p = 0.300$).

When the international and Hungarian students are compared, there is no significant difference between the Excel results ($Z(E) = -1.036$, $p = 0.219$). However, there is a significant difference between the paper (P15) results ($Z(P15) = -3.056$, $p = 0.002$); in general, the paper results of the Hungarian students are better than of the international students.

### 4.4. Comparison of self-assessment and results

To recognize students' attitude towards spreadsheeting, we also recorded their self-assessment values, as mentioned above, on a 0–100 scale. These values were taken four times (SAV1–SAV4); however, for the present study, the first evaluation is compared to the solutions on paper and in Excel, respectively.

The mean and standard deviation results of the self-assessment values are presented in Table 8. It is found with the Shapiro-Wilk Test that the distribution of SAV1 is not normal (Shapiro-Wilk Statistic = 0.977, $p < 0.006$; Shapiro-Wilk Statistic = 0.951, $p < 0.001$; Shapiro-Wilk Statistic = 0.888, $p < 0.001$ respectively)). Figure 10 shows the histogram of SAV1.

|   | Mean | Std |
|---|---|---|
| SAV1 | 47.33 | 21.35 |
| paper (P15) | 35.44 | 21.35 |
| Excel | 35.13 | 24.41 |

Table 8. Students' first self-assessment values (SAV1) and their test results (paper and Excel results copied from Table 3).

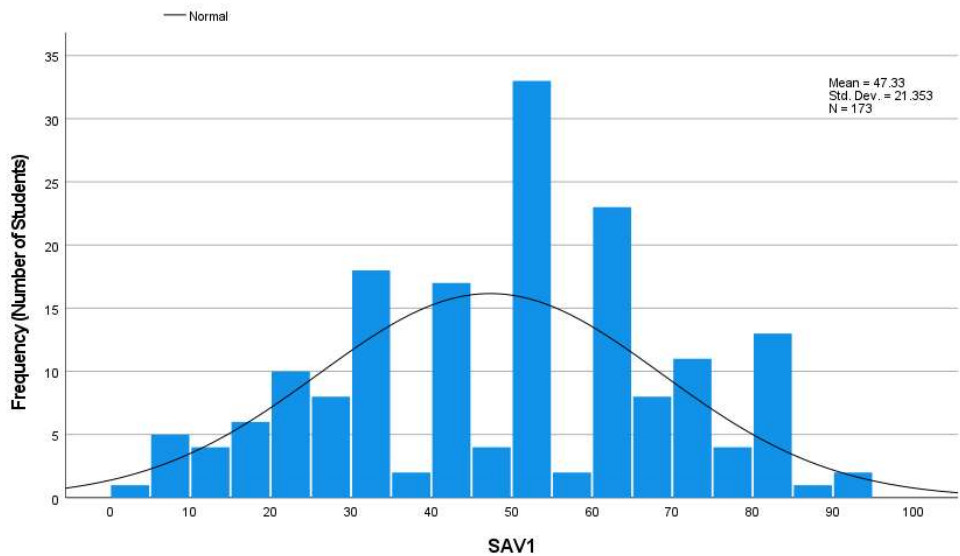

Figure 10. The distribution of SAV1 values.

We compare students' self-assessment spreadsheet knowledge, skills, and competencies compared to their results on paper and in Excel (SAV1 vs. P15 and SAV1 vs Excel). Testing for mean differences, the Wilcoxon Signed Ranks Tests revealed significant negative differences (ranks) between SAV1 and the paper (P15) results (z statistics = −5.161, p = < 0.001) and between SAV1 and the Excel results (z statistics = −5.133, p = < 0.001). We can accept the second hypothesis [H1]. Compared to their paper and Excel scores, students are less accurate in assessing their spreadsheet knowledge, skills, and competences. However, for further results and proofs, we need additional statistical analyses.

|  | P17 | P15 | E |
|---|---|---|---|
| SAV1 | 0.1452* | 0.1252 | 0.1776** |

Note: ***p < 0.001, **p < 0.05, *p < 0.01

Table 9. The correlation coefficient compares the SAV1 self-assessment values and results on paper for Tasks 1–7 (P17) and Tasks 1–5 (P15) in Excel.

The correlation coefficients reveal a weak connection between the self-assessment values and the results (Table 9). These findings prove that self-assessment values are unreliable to predict end-users' spreadsheet knowledge. The Dunning-Kruger effect works (Kruger & Dunning, 1999; Csernoch et al., 2021; Gibbs et al., 2011, 2014, 2017; Kun et al., 2023). However, our study also revealed that when real problems are presented, these values are significantly affected.

## 5. CONCLUSIONS

The present study provides the details of the preliminary results of a large-scale research to test tertiary students' spreadsheet skills, competences, problem-solving strategies, and confidence in their knowledge. The research aims to test students' spreadsheet knowledge both on paper and in Excel and compare these results with their self-assessment values, which were taken four times during the testing.

To collect data, students solve five tasks both on paper and in Excel and evaluate two formulas on paper. In the digital solution, a prepared spreadsheet table is presented, and the same five tasks should be solved with formulas. The evaluation of the students' results is based on the item points assigned to the smallest recognizable pieces of formulas and sentences. In addition to comparing the scores, the digital test also records the students' activities. Two log files and their spreadsheet solutions are created. The log files allow us to reveal the students' problem-solving and searching strategies, as well as the time spent on the tasks.

The results indicate that based on the students' previous studies and experience, their first self-evaluation values show only a weak correlation with scores, and in general these values are much higher than the results in the test. It is also revealed, on average, it takes at least twice as long to achieve the same results in Excel as on paper. We can conclude that our findings do not support Prensky's digital native-immigrant idea but are in line with those who question Prensky's categorization and arguments.

Further analysis is needed to find out more about students' problem-solving strategies and the discrepancy between self-assessment scores and actual performance. The data recorded in the log files and the four evaluation values collected during the testing process would allow us to identify further characteristics of students and end-users considering their activities and behaviours during digital problem-solving. These results would help us to build novel teaching-learning strategies that are more effective and efficient than the now wildly applied but questioned.

**REFERENCES**

Abonyi-Tóth, A., Farkas, C., Reményi, Z., & Jeneiné Horváth, K. (2021). *Digital Culture 10. In Hungarian: Digitális kultúra 10.* Oktatási Hivatal. https://www.tankonyvkatalogus.hu/pdf/OH-DIG10TA__teljes.pdf

Abonyi-Tóth, A., Farkas, C., Fodor, Z., Jeneiné Horváth, K., Reményi, Z., Siegler, G., & Varga, P. (2022). *Digital Culture 11., In Hungarian: Digitális kultúra 11.* Oktatási Hivatal. https://www.tankonyvkatalogus.hu/pdf/OH-DIG11TA__teljes.pdf

Biro, P., & Csernoch, M. (2014). Deep and surface metacognitive processes in non-traditional programming tasks. *2014 5th IEEE Conference on Cognitive Infocommunications (CogInfoCom)*. https://doi.org/10.1109/coginfocom.2014.7020507

Csapó, G., Csernoch, M., & Abari, K. (2019). Sprego: Case study on the effectiveness of teaching spreadsheet management with schema construction. *Education and Information Technologies*, *25*(3), 1585–1605. https://doi.org/10.1007/s10639-019-10024-2

Csapó, G., Sebestyén, K., Csernoch, M., & Abari, K. (2020). Case study: Developing long-term knowledge with Sprego. *Education and Information Technologies*, *26*(1), 965–982. https://doi.org/10.1007/s10639-020-10295-0

Csernoch, M. (2017). Thinking fast and slow in computer problem solving. *Journal of Software Engineering and Applications*, *10*(01), 11–40. https://doi.org/10.4236/jsea.2017.101002

Csernoch, M., & Biró, P. (2015). Computer Problem Solving. In Hungarian: Számítógépes problémamegoldás. *Tudományos és Műszaki Tájékoztatás*, *62*(3), 86–94. https://epa.oszk.hu/03000/03071/00084/pdf/EPA03071_tmt_2015_03_086-094.pdf. (Accessed: 03.11.2024)

Csernoch, M., Biró, P., & Máth, J. (2021). Developing computational thinking skills with algorithm-driven spreadsheeting. *IEEE Access*, *9*, 153943–153959. https://doi.org/10.1109/access.2021.3126757

Csernoch, M., Máth, J., & Nagy, T. (2022). The interpretation of graphical information in word processing. *Entropy*, *24*(10), 1492. https://doi.org/10.3390/e24101492

Csernoch, M., Nagy, K., & Nagy, T. (2023). The entropy of digital texts—the mathematical background of correctness. *Entropy*, *25*(2), 302. https://doi.org/10.3390/e25020302

Csernoch, M., Nagy, T., Nagy, K., Csernoch, J., & Hannusch, C. (2024). Human-centered digital sustainability: Handling enumerated lists in digital texts. *IEEE Access*, *12*, 30544–30561. https://doi.org/10.1109/access.2024.3369587

Gibbs, S., McKinnon, A., & Steel, G. (2014). *Are workplace end-user computing skills at a desirable level? A New Zealand perspective*. AIS Electronic Library (aisel) - AMCIS 2014 proceedings: Are workplace end-user computing skills at a desirable level? A New Zealand perspective. https://aisel.aisnet.org/amcis2014/EndUserIS/GeneralPresentations/5/

Gibbs, S., Moore, K., Steel, G., & McKinnon, A. (2017). The Dunning-Kruger effect in a Workplace Computing Setting. *Computers in Human Behavior*, *72*, 589–595. https://doi.org/10.1016/j.chb.2016.12.084

Gibbs, S. F., Steel, G., & Kuiper, A. (2011). Expectations of competency: The mismatch between employers' and graduates' views of end-user computing skills requirements in the Workplace. *Journal of Information Technology Education: Research*, *10*, 371–382. https://doi.org/10.28945/1531

ICDL (2023). *ICDL Workforce SPREADSHEETS Syllabus 6.0*. https://icdl.org/app/uploads/2024/01/ICDL-Computer-Online-Essentials-Syllabus-1.0.pdf

Kahneman, D. (2011). *Thinking, fast and slow*. Penguin Books, UK.

Kruger, J., & Dunning, D. (1999). Unskilled and unaware of it: How Difficulties in Recognizing One's Own Incompetence Lead to Inflated Self-Assessments. *Journal of Personality and Social Psychology*, *77*(6), 1121–1134. https://doi.org/10.1037/0022-3514.77.6.1121

Kun, A. I., Juhász, C., & Farkas, J. (2023). Dunning–kruger effect in knowledge management examination of BSC Level Business Students. *International Journal of Engineering and Management Sciences*, *8*(1), 14–21. https://doi.org/10.21791/ijems.2023.1.3.

László, V. C., Nagy, K., & Csernoch, M. (2022). 15. Informatika Szakmódszertani Konferencia. In P. Szlávi & L. Zsakó (Eds.), *INFODIDACT'2022* (pp. 127–140). Zamárdi; Webdidaktika Alapítvány. Retrieved from https://people.inf.elte.hu/szlavi/InfoDidact22/Infodidact2022.pdf.

Máté, D., & Darabos, É. (2017). Measuring the accuracy of self-assessment among undergraduate students in higher education to enhance competitiveness. *Journal of Competitiveness*, *9*(2), 78–92. https://doi.org/10.7441/joc.2017.02.06

MS (2024). Mo-210: Microsoft Excel (Microsoft 365 apps). https://query.prod.cms.rt.microsoft.com/cms/api/am/binary/RE5axcj

Nagy, K., & Csernoch, M. (2023). Pre-testing erroneous text-based documents: Logging end-user activities. *Frontiers in Education*, *7*. https://doi.org/10.3389/feduc.2022.958635

Nagy, T., Csernoch, M., & Biró, P. (2021). The comparison of students' self-assessment, gender, and programming-oriented spreadsheet skills. *Education Sciences*, *11*(10), 590. https://doi.org/10.3390/educsci11100590

OH (2004, June 1). *Próbaérettségi 2004 - választható tárgyak*. A 2004. évi kétszintű próbaérettségi írásbeli feladatsorai és megoldásai a választható tárgyakból, közép- és emelt szinten. https://www.oktatas.hu/bin/content/dload/erettsegi/probaerettsegi_2004/info_em_flap.zip

OH (2020). *Közismereti érettségi vizsgatárgyak 2024. május-júniusi vizsgaidőszaktól érvényes vizsgakövetelményei (2020-as Nat-ra épülő vizsgakövetelmények)*. Részletes vizsgakövetelmények és vizsgaleírások. https://www.oktatas.hu/pub_bin/dload/kozoktatas/erettsegi/vizsgakovetelmenyek2024/dig_kult_2024_e.pdf

Panko, R. R. (2013). The Cognitive Science of spreadsheet errors: Why thinking is bad. *2013 46th Hawaii International Conference on System Sciences*. https://doi.org/10.1109/hicss.2013.513

Panko, R. R. (2015). What We Don't Know About Spreadsheet Errors Today: The Facts, Why We Don't Believe Them, and What We Need to Do. In S. Thorne & G. Croll (Eds.), *Proceedings of the EuSpRIG 2015 Conference "Spreadsheet Risk Management"* (pp. 73–87). essay, European Spreadsheet Risks Interest Group. ISBN: 978-1-905404-52-0

Prensky, M. (2001a). Digital Natives, digital immigrants. *On the Horizon*, *9*(5), 1–6. https://doi.org/10.1108/10748120110424816

Prensky, M. (2001b). Digital Natives, Digital Immigrants Part 2: Do they really think differently? *On the Horizon*, *9*(6), 1–6. https://doi.org/10.1108/10748120110424843

Sweller, J., Ayres, P., & Kalyuga, S. (2011). *Cognitive load theory*. Springer New York.

Varga, P., Jeneiné Horváth, K., Reményi, Z., Farkas, C., Takács, I., Siegler, G., & Abonyi-Tóth, A. (2020). *Digital Culture 9. In Hungarian: Digitális kultúra 9*. Oktatási Hivatal, Budapest. https://www.tankonyvkatalogus.hu/site/kiadvany/OH-DIG09TA